\documentclass[aps,prfluids]{revtex4-2}
\usepackage{amssymb}
\usepackage{graphicx}
\usepackage{bm}
\usepackage{physics}

\begin{document}
\title{On force balance in Brinkman fluids under confinement} 

\author{Abdallah Daddi-Moussa-Ider}
\affiliation{School of Mathematics and Statistics, The Open University, Walton Hall, Milton Keynes MK7 6AA, United Kingdom}
\author{Andrej Vilfan}
\email{andrej.vilfan@ijs.si}
\affiliation{Jo\v{z}ef Stefan Institute, 1000 Ljubljana, Slovenia}
\date{\today}

\begin{abstract}
A point force acting on a Brinkman fluid in confinement is always counterbalanced by the force on the porous medium, the force on the walls and the stress at open boundaries. We discuss the distribution of those forces in different geometries: a long pipe, a medium with a single no-slip planar boundary, a porous sphere with an open boundary and a porous sphere with a no-slip wall. We determine the forces using the Lorentz reciprocal theorem and additionally validate the results with explicit analytical flow solutions. We discuss the relevance of our findings for cellular processes such as cytoplasmic streaming and centrosome positioning.
\end{abstract}

\maketitle

\section{Introduction}

\begin{figure}
\centering
\includegraphics[width=\linewidth]{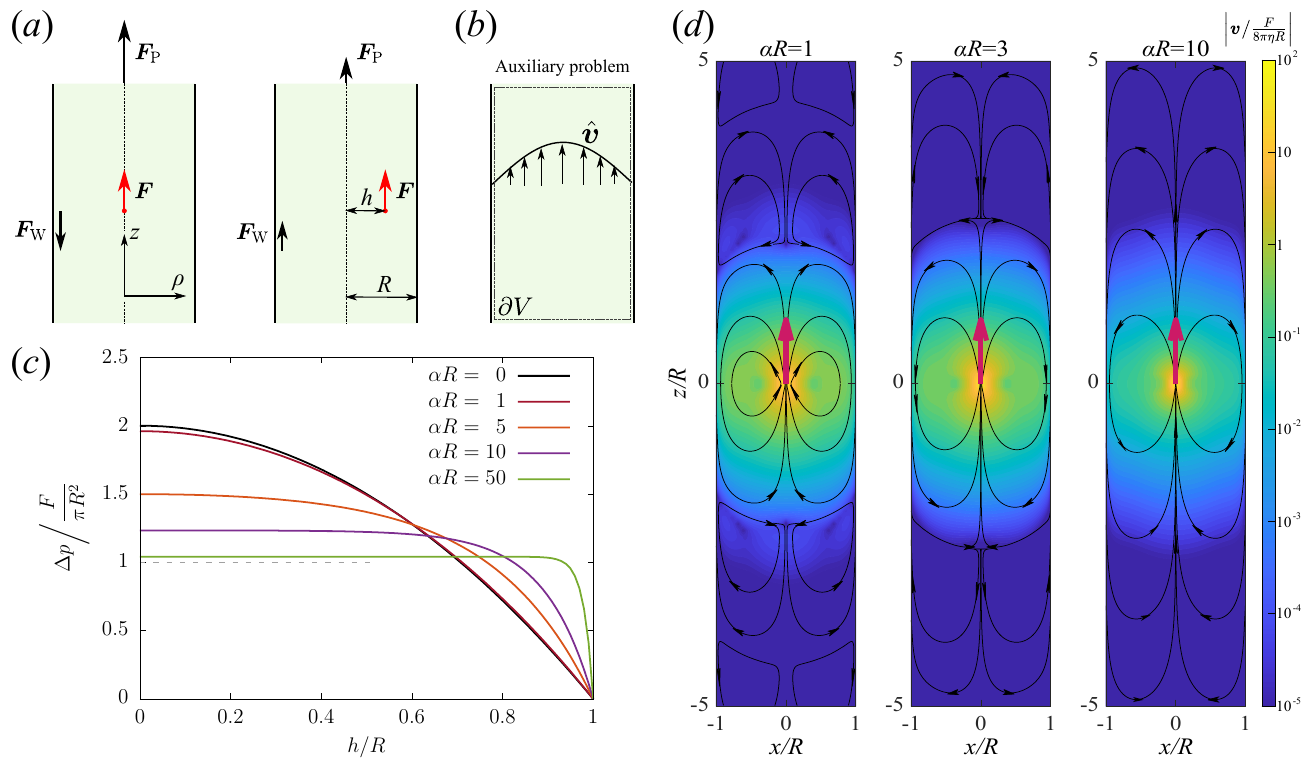}
\caption{Point force in a pipe. (a)~Force balance: the external force $F$ is distributed to the force on the wall $F_\text{W}$ and the force due to pressure difference $F_\text{P}=\pi R^2 \Delta p$. The direction of the force on the walls changes with the radial position of the force. (b)~Pressure driven flow through a pipe with a Brinkman fluid is used as an auxiliary problem when applying the reciprocal theorem. (c)~Induced pressure difference as a function of the distance $h$ of the force from the tube's axis, as given by Eq.~\eqref{eq:Delp_tube}. The dashed line represents the limit $\alpha \to \infty$. (d)~Streamlines around a point force acting at the centre of the channel ($h=0$). At small values of $\alpha R$, the flow shows a series of viscous tubular eddies which disappear at around $\alpha R\approx 4.7$.
}
\label{fig:tube}
\end{figure}

In the absence of of inertia, fluids at low Reynolds numbers are always in a state of force balance: the sum of volume forces, surface forces and the stress integrated over a closed surface always equals zero. In confined geometries, the force balance can take a surprising form. A notable example is the flow due to a point force inside a long pipe with a no-slip boundary \citep{Brenner1959,Liron.Shahar1978}. The external force leads to a pressure difference between the two ends of the pipe (Fig.~\ref{fig:tube}a). Depending on the radial position of the point force, the pressure difference, multiplied by the cross-section area of the pipe, can exceed the original force by a factor of 2 \citep{Pliskin.Brenner1963,MONDY.BRENNER1997}. The remainder of the force is contributed by the force on the wall of the pipe, which changes its sign at a certain position of the applied force. A related result was found for a Stokeslet between two parallel plates \citep{Skultety.Morozov2020}. 

The Brinkman equation~\citep{brinkman1949A, brinkman1949B}
\begin{equation}
\label{eq:brinkman}
\eta (\Delta \bm{v}-\alpha^2 \bm{v})-\nabla p+\bm{f}=\bm{0}\,,
\end{equation}
along with the incompressibility condition $\nabla \cdot \bm{v}=0$,
describes the flow of a fluid through a porous medium.  
Here, \(\eta\) is the dynamic viscosity of the Newtonian fluid and \(\alpha^2\) represents the impermeability of the porous medium, with dimensions of (length)\(^{-2}\).
The density of external forces acting on the fluid is denoted by $\bm{f}$. 
Equivalently, it can also be expressed as a continuity equation for the stress tensor $\bm{\sigma}=-p \bm{I}+ \eta (\nabla \bm{v} + (\nabla \bm{v})^T)$:
\begin{equation}
\label{eq:brinkman_sigma}
\nabla\cdot \boldsymbol{\sigma}-\eta \alpha^2\bm{v}+\bm{f}=\bm{0} \,.
\end{equation}
The force balance therefore needs to take into account the momentum exchange between the fluid and the medium. The force balance equation on a volume of fluid $\mathcal V$ is obtained by applying the divergence theorem to Eq.~\eqref{eq:brinkman_sigma} and reads
\begin{equation}
     \oint_{\partial \mathcal V} \bm{\sigma }\cdot \bm{n} \, \dd S + \int_{\mathcal V} (\bm{f} -\eta \alpha^2 \bm{v}) \, \dd V =\bm{0}\,,
\end{equation}
with $\bm{n}$ as a unit normal vector pointing outward. In this paper, we discuss the cumulative force balance in Brinkman flows under different types of confinement. We discuss the application of the results to different kinds of cytoplasmic flows.

\section{Examples of force balance}
The discussion of specific cases follows the level of confinement. We begin with the flow induced by a point force in an infinite pipe, akin to the solution by \citet{Liron.Shahar1978}, but with a Brinkman medium instead. We continue with the point force in the presence of a wall, extending the classical solution of \citet{blake71}, with a force balance specifically discussed by \citet{Skultety.Morozov2020}. Finally, we proceed with a force inside a porous sphere with open boundaries (i.e., surrounded by free fluid), as well as no-slip spherical confinement. All solutions are determined by the application of the Lorentz reciprocal theorem, which greatly simplifies the calculation of cumulative forces. We also validate and illustrate the results with explicit exact solutions for some cases. 

\subsection{Material force in confined flows}
\label{sec:confined}
We begin with a note on the material force, i.e., the force the fluid exerts on the porous medium, 
\begin{equation}
\bm{F}_\text{M}=\int_{\mathcal{V}} \eta \alpha^2 
\bm{v} \, \dd V
\end{equation}
in a confined flow with zero normal velocity, \( \bm{v}\cdot \bm{n}=0\) at the outer boundary $\partial \mathcal V$. By applying the divergence theorem to the quantity $(\bm{e}\cdot \bm{x}) \bm{v}$ with an arbitrary unit vector $\bm{e}$ and noting that $\nabla \cdot ((\bm{e}\cdot \bm{x}) \bm{v})= \bm{e}\cdot \bm{v}$ for $\nabla \cdot \bm{v}=0$, we obtain
\begin{equation}
  \label{eq:confinement}
    \bm{e}\cdot \int_{\mathcal{V}} 
\bm{v} \, \dd V = \oint_{\partial \mathcal{V}} (\bm{e}\cdot \bm{x}) \bm{v} \cdot \bm{n} \, \dd S =0\,.
\end{equation}
Because the equality holds for any $\bm{e}$, it follows that
\begin{equation}
\int_{\mathcal{V}} 
\bm{v} \, \dd V =\bm{0}
\end{equation}
for any confined flow, and therefore the total force acting on the porous material vanishes, $\bm{F}_\text{M}=\bm{0}$. 

\subsection{Point force in a pipe}
\label{sec:pipe}

The first example consists of a point force $F$ inside an infinitely long pipe with radius $R$ that is filled with a porous medium (Fig.~\ref{fig:tube}a). The force is acting in axial direction at distance $h$ from the axis of the pipe. We disregard any radial components of the force $F$, because it is clear from symmetry arguments that they will act on the wall of the tube alone, without contributing to the long-range pressure difference.

In order to apply the Lorentz reciprocal theorem, we first solve the auxiliary problem, which is a pressure driven flow through a pipe containing a Brinkman medium (Fig.~\ref{fig:tube}(b)). The fluid is therefore described by the Brinkman equation~\eqref{eq:brinkman}, along with the condition of incompressibility $\nabla \cdot \bm{\hat v}=0$ (the hatted quantities are used to distinguish the auxiliary from the main problem.) The pipe is oriented along the $z$-axis. At the walls the no-slip condition $\bm{\hat v}=0$ applies and the whole tube is subject to a negative pressure gradient, $G=-\mathrm{d} \hat p / \mathrm{d} z > 0$. We write the Brinkman equation in cylindrical coordinates
\begin{equation}
    \frac{1}{\rho} \frac{\mathrm{d} }{\mathrm{d} \rho} 
    \left( \rho \, \frac{\mathrm{d} \hat{v}_z}{\mathrm{d} \rho} \right) - \alpha^2 \hat{v}_z + \frac{G}{\eta}
    = 0 \, . \label{eq:brinkman-tube}
\end{equation}
The boundary condition states \( \hat{v}_z(\rho = R) = 0 \).
The solution for the axial velocity is 
\begin{equation}
    \hat{v}_z = \frac{G}{\alpha^2 \eta}
    \left( 1 - \frac{I_0(\alpha \rho)}{ I_0(\alpha R) } \right) \, , 
\end{equation}
with $I_n(x)$ denoting the $n$th-order modified Bessel function of the first kind~\citep{abramowitz72}. The volume flow rate through the pipe is 
\begin{equation}
    \hat Q = \int_0^R \hat v_z (\rho) \, 2\pi\rho \, \dd\rho 
    = \frac{\pi R^2 G}{\alpha^2 \eta} \frac{I_2(\alpha R)}{ I_0(\alpha R) } \;.
\end{equation}

In the limit $\alpha \to 0$, corresponding to the Stokes flow regime, we can use the series expansion $I_0(x)= 1+ x^2/4 + \mathcal{O}(x^4)$, $I_2(x)= x^2/8 + \mathcal{O}(x^4)$ and recover the classical parabolic velocity profile of the Hagen–Poiseuille flow with \(\hat v _z=G (R^2-r^2) /(4\eta)\) and \(\hat Q=\pi R^4 G/(8\eta)\).

We now apply the Lorentz reciprocal theorem (derived for Brinkman fluids in Appendix~\ref{appendix:lrt}) with the pressure driven flow as auxiliary problem. 
The reciprocal theorem states
\begin{equation}
 \int_{\mathcal \partial \mathcal V} \bm{v}\cdot \hat {\boldsymbol\sigma} \cdot \bm{n} \, \dd S +
 \int_{\mathcal V} \bm{v} \cdot \bm{\hat f} \, \dd V=
    \int_{\mathcal \partial \mathcal V} \bm{\hat v}\cdot \boldsymbol\sigma \cdot \bm{n} \, \dd S +
    \int_{\mathcal V}  \bm{\hat v} \cdot \bm{f} \, \dd V 
\,.   
\end{equation}
The integration surface $\partial \mathcal V$ runs along the wall of the pipe and is closed at $z=\pm \infty$ (Fig.~\ref{fig:tube}b). 
Because both velocities are zero at the walls, all surface integrals vanish there. At the ends of the pipe, the left surface integral vanishes because $\bm{v}\to 0$, but the right gives $-\hat Q \Delta p$, where $\Delta p=p(z\to \infty)-p(z\to -\infty)$ is the pressure difference buildup in the main problem. We know that the pressure at both ends needs to be constant across the tube because of the vanishing velocity. The left volume integral also vanishes because $\bm{\hat f}=\bm{0}$ by definition. This leads to a relationship for the pressure difference
\begin{equation}
    \Delta p=\frac{F \hat v_z(h)}{\hat Q}= \frac{F}{\pi R^2} \frac {I_0(\alpha R)-I_0(\alpha h)}{I_2(\alpha R)}\;.
    \label{eq:Delp_tube}
\end{equation}
In the limit $\alpha\to 0$, the relation simplifies to $\Delta p=2 F (1-h^2/R^2)/(\pi R^2)$, in agreement with the exact flow solution \citep{Brenner1959,Liron.Shahar1978} and also the lubrication approximation by \cite{Navardi.Bhattacharya2010}. Note that the reciprocal theorem allowed us to obtain this result with a much shorter and simpler derivation compared to the explicit solutions in the references above.

The resulting radial dependence is shown in Fig.~\ref{fig:tube}(c). Whereas the Brinkman medium always reduces the pressure if the force is near the axis of the channel, the opposite is the case in the proximity of the wall, where an increased value of~$\alpha$ leads to a higher $\Delta p$. 

Because of the rapid fall-off of the velocity with the distance along the pipe, the pipe can be effectively seen as confined, and the force on the porous material is zero. Force balance therefore implies that the force exerted by the fluid on the walls is 
\begin{equation}
F_\text{W}= F-\pi R^2 \Delta p=F\left( 1 - \frac {I_0(\alpha R)-I_0(\alpha h)}{I_2(\alpha R)} \right). \label{eq:FW_pipe}
\end{equation}
As in the Stokes fluid, the force on the wall changes direction: it is parallel to $F$ when the external force acts close to the wall, but antiparallel when it is in the centre of the pipe.

Although the we only consider an infinitely long tube here, the solution for a length that is finite, but still much larger than the diameter, can easily be obtained by superposition. Another possible extension is the mobility matrix of two particles in the pipe. For Stokes fluids, it has been shown that the off-diagonal elements expressing hydrodynamic interactions depend on the length of the tube, but not on the distance between the particles \citep{Misiunas.Keyser2015}.

\subsection{Point force in the proximity of a plane}
\begin{figure}
\centering
\includegraphics[width=0.9\linewidth]{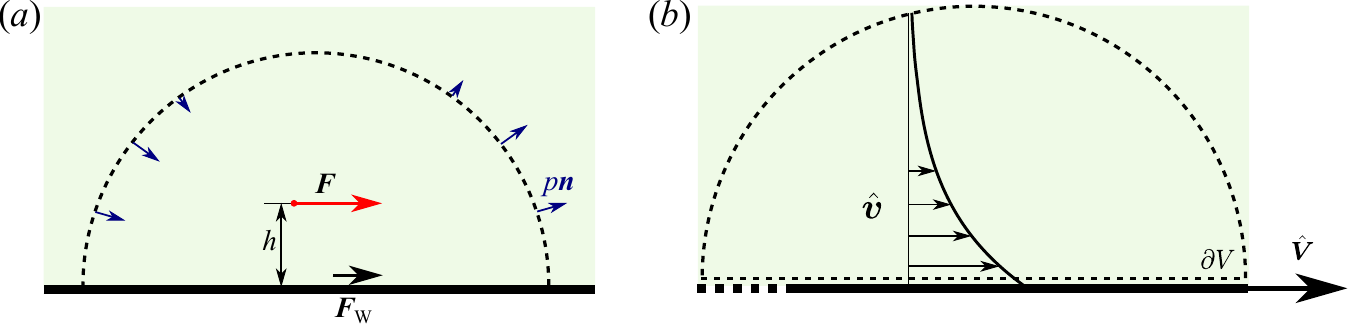}
\caption{
(a)~Point force $\bm F$ on the Brinkman fluid in the presence of a no-slip wall. The force is distributed over the force on the wall $\bm{F}_\text{W}$, the force on the porous medium $\bm{F}_\text{M}$ and force due to the pressure difference. (b)~In the auxiliary problem, the wall itself is moving with the velocity $\hat{\bm{V}}$. 
}
\label{fig:plane}
\end{figure}

We now study the force exerted on an infinite plane by a force acting parallel to it at some distance $h$ (Fig.~\ref{fig:plane}a). The auxiliary problem consists of the plane itself moving tangentially with velocity $\hat{\bm{V}}$ while the porous medium itself does not move (Fig.~\ref{fig:plane}b). The Brinkman equation for the auxiliary problem reduces to
\begin{equation}
    (\partial_z^2 -\alpha^2) \hat{\bm{v}}=\bm{0} \, , 
\end{equation}
with the boundary condition $\hat{\bm{v}}=\hat{\bm{V}}$ at $z=0$ and is solved as
\begin{equation}
    \hat{\bm{v}}=\hat{\bm{V}} e^{-\alpha z}\,.
\end{equation}

We now apply the Lorentz reciprocal theorem to the half-space containing the Brinkman fluid. The integration surface contains the plane and the upper hemisphere at infinite distance (dashed line in Fig.~\ref{fig:plane}b). Because both $\hat{\bm v}$ and $\bm v$ converge to 0 sufficiently fast at large distances, the surface integrals over the hemisphere vanish. The reciprocal theorem then states that
\begin{equation}
 \int \hat{\bm V} \cdot \boldsymbol{\sigma}\cdot \bm{e}_z \, \dd S  -  \hat {\bm{v}} \cdot \bm F  = 0
\end{equation}
leading to \( \hat{\bm V} \cdot  \bm{F}_\text{W}= \hat {\bm{v}}\cdot \bm F \). The horizontal force exerted by the fluid on the plane is therefore
\begin{equation}
  F_\text{W} = F e^{-\alpha z}\,. \label{eq:force-wall}
\end{equation}
The long-range nature of the flow and pressure (calculated explicitly in Eq.~\eqref{eq:plane-p}) has the consequence that the integral $\int p \bm{n} \,\dd S$ does not have a defined limit when the boundary is infinitely far from the force. Therefore, it is not possible to state which part of the force is carried by pressure and which part by the porous medium, unlike the force on the wall. However, the force due to the pressure vanishes in the Stokes limit $\alpha\to 0$, as already noted by \citet{blake71}. A related problem appears in a Stokes fluid between two plates, which becomes related to a 2D Brinkman fluid in the far-field regime. Although the pressure contribution to the horizontal force balance was determined by \cite{Skultety.Morozov2020}, we note that the exact result depends on the way how the integration area approaches an infinite plane and is therefore not uniquely defined.

\subsection{Point force inside a sphere with open boundaries}
\label{sec:2.4}

\begin{figure}
\centering
\includegraphics[width=\linewidth]{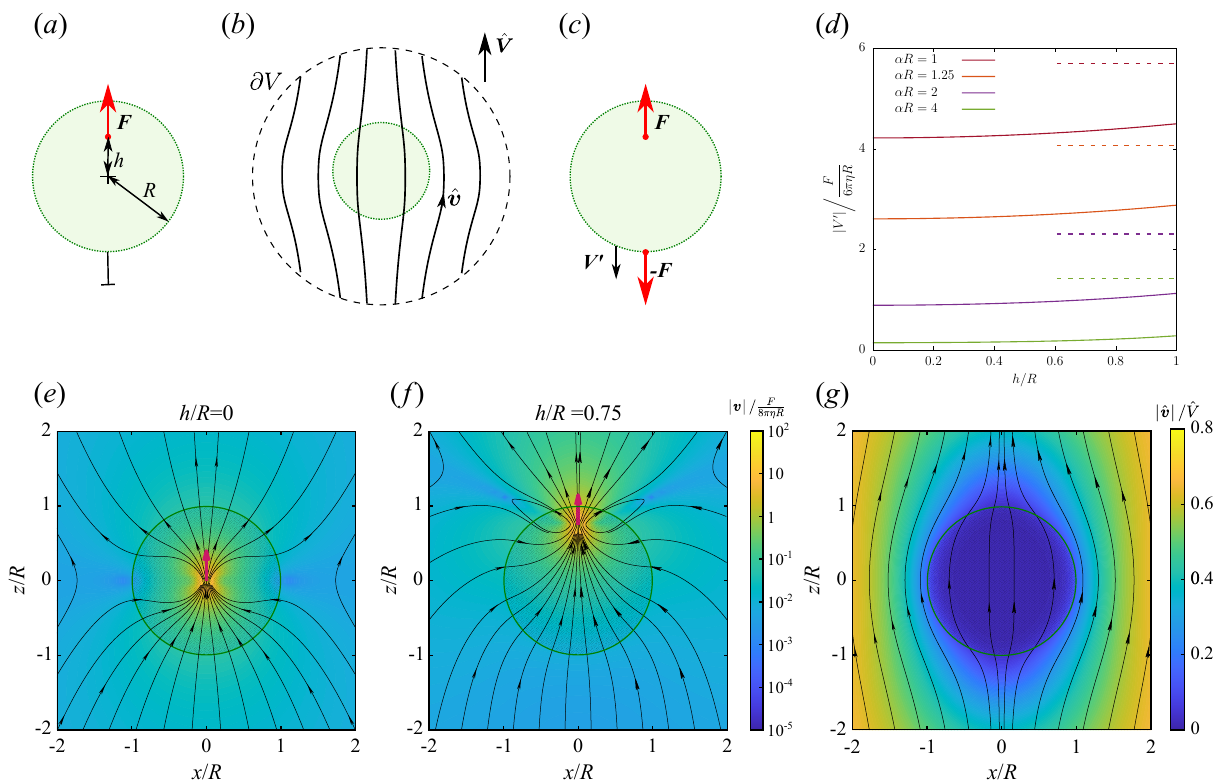}
    \caption{(a)~The force $\bm{F}$ is acting axisymmetrically on the Brinkman fluid inside a porous sphere at a fixed position in laboratory frame and embedded in a Stokes fluid. (b)~The same sphere, subject to an external flow with velocity $\hat {\bm{V}}$, is used as auxiliary problem. (c)~The porous sphere is free to move, and subject to a pair of opposite forces, one acting on the Brinkman fluid and one on the sphere. (d)~The velocity $\bm{V}^\prime$ of the sphere shown in panel (c)~as a function of the force position. The horizontal dashed lines indicate the  velocity of the porous sphere pulled by an external force of magnitude $\bm{F}$. (e,f) Flow in and around the fixed porous sphere for two positions of the external point force. (g) Flow in the auxiliary problem. All streamline plots are computed with the value $\alpha R=10$.}
    \label{fig:sphere}
\end{figure}

In this section, we consider an external point force acting on the fluid inside a porous sphere, immersed in a Stokes fluid of the same viscosity (Fig.~\ref{fig:sphere}(a)). The porous medium itself is fixed in the laboratory frame.

For the auxiliary problem (Fig.~\ref{fig:sphere}(b)), we choose the same sphere in an external flow such that $\bm{\hat v}(r\to \infty)=\bm{\hat V}$. A solution for the flow past a porous sphere can be found in \cite{Felderhof1975}. The drag coefficient is 
\begin{equation}
  \label{eq:drag}
  \Gamma = 4 \pi \eta R \, (\alpha R)^2 \, 
  \frac{  \phi(\alpha R) }{\phi(\alpha R)+2} \, ,
\end{equation}
where we introduced the function
\begin{equation}
    \phi(x) = \frac{3}{x^2} \left( 1-\frac{\tanh x}{x} \right) ,
\end{equation}
which takes values between $1$ when \( x = 0 \) and $0$ when \( x \to \infty \).
The radial flow velocity along the symmetry axis, at distance $r$ from the centre of the sphere, is \citep{Felderhof1975}
\begin{equation}
    \bm{\hat v}=\bm{\hat V} \, \frac{   
     \phi(\alpha R)  + 2\,\phi(\alpha r)\cosh(\alpha r)/\cosh(\alpha R)  } { \phi(\alpha R) +2} .
\end{equation}
An example of the complete flow profile is shown in Fig.~\ref{fig:sphere}(g).

For simplicity, we restrict ourselves the case where the external force is axisymmetric, acting radially with respect to the centre of the sphere. 
We apply the reciprocal theorem to the whole space, closed at infinity (Fig.~\ref{fig:sphere}(b)). At the integration surface, we therefore have $\bm{v}\to 0$ and $\oint \bm{\sigma}\cdot \bm{n} \, \dd S = -\bm{F}^\infty$, the force transmitted to the fluid. The velocity of the auxiliary problem is $\bm{\hat v}=\bm{\hat V}$. The reciprocal theorem then reads
\begin{equation}
  \bm{F}\cdot \bm{\hat v}(h) -  \bm{F}^\infty \cdot \bm{\hat V}=0\,.
\end{equation}
The resulting force is
\begin{equation}
    \bm{F}^\infty =\bm{F} \, \frac{   
     \phi(\alpha R)  + 2\,\phi(\alpha h)\cosh(\alpha h)/\cosh(\alpha R)  } {\phi(\alpha R) +2} . \label{eq:force-sphere}
\end{equation}
Likewise, the force exerted on the porous sphere is
\begin{equation}
\bm{F}_\text{M}=\bm{F}-\bm{F}^\infty=2F \, \frac{1- \phi(\alpha h) \cosh(\alpha h)/\cosh(\alpha R)  } {\phi(\alpha R) +2 }\,.
\end{equation}

In the above calculation, the force is external in its nature and the porous sphere is fixed in space. A more interesting example, however, is the situation where the force acts between the porous medium and the fluid, i.e., if the force~$\bm F$ acts on the fluid and simultaneously the force $-\bm F$ on the medium (Fig.~\ref{fig:sphere}(c)). This model can describe a motor protein pulling a vesicle through a dense filament network. For some continuous force distributions, a similar problem, albeit with a different boundary condition, has been solved by \citet{Kree.Zippelius2018}. We solve the problem for a point force in two steps: we initially keep the porous sphere fixed in space, and later determine its velocity under the force-free condition. For a clamped sphere, the total force on the medium is
\begin{equation}
  \bm{F}_\text{M}'=-\bm{F}+\bm{F}_\text{M}=-\bm{F}^\infty\,.
\end{equation}
If the sphere is free to move, such that the only forces acting on it are hydrodynamic forces and the reaction $-\bm F$, its velocity will be
\begin{equation}
\label{eq:vprime_open}
  \bm{V}' = \bm{F}_\text{M}'/\Gamma = 
  -\frac{F}{4\pi\eta R (\alpha R)^2 } \left(1+2 \,
  \frac{\phi(\alpha h) \cosh(\alpha h)}{\phi(\alpha R) \cosh(\alpha R) } \right) \,.
\end{equation}
The velocity as a function of the position and the constant $\alpha$ is shown in Fig.~\ref{fig:sphere}(d). The velocity always lies below what would be expected when the medium was pulled by an external force of magnitude $-\bm F$. 
The velocity magnitude has a minimum at the centre of the sphere and increases weakly and monotonically with $h$, attaining its maximum value at $h=R$. As $\alpha R$ decreases, the velocities increase, where $V' \sim \alpha^{-2}$ around $\alpha=0$.

\subsection{Point force inside a sphere with no-slip confinement}
\label{sec:2.5}
\begin{figure}
    \centering
\includegraphics[width=0.6\textwidth]{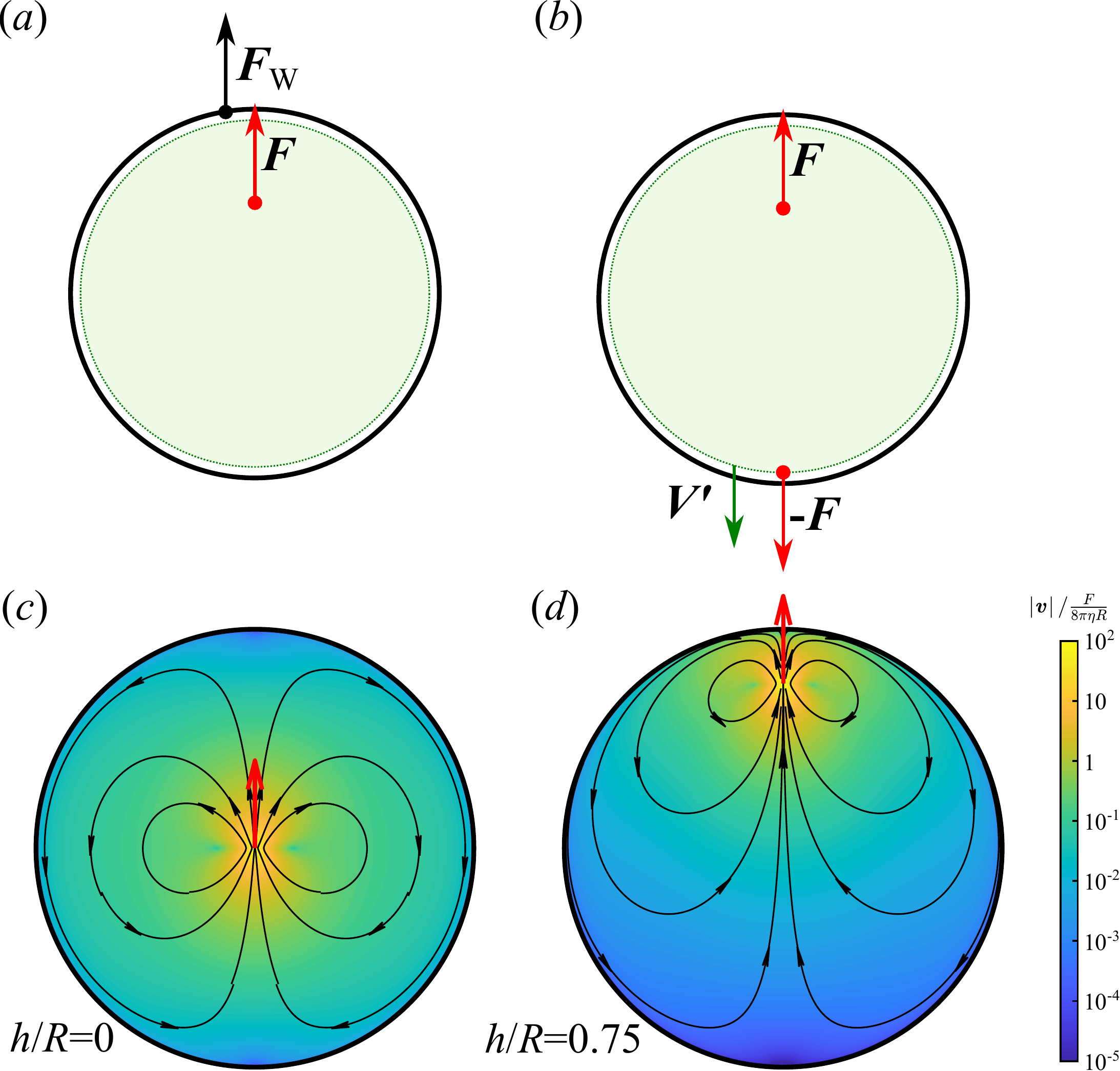}
    \caption{(a)~Point force acting on the fluid in a porous sphere, confined by a no-slip boundary. All the force is transmitted to the wall, $\bm{F}_\text{W}=\bm{F}$. (b)~A pair of forces acting between the fluid and the porous sphere. The sphere is enclosed by the no-slip boundary at close distance, but still free to move (with velocity $\bm{V}'$). (c,d) The flow velocity induced by a point force located at distance $h/R=0$ and $h/R=0.75$, both for \(\alpha R = 10\).}
    \label{fig:noslip}
\end{figure}

The final example also concerns a point force in a porous sphere, which is now subject to additional confinement by a spherical wall with no-slip conditions. The sphere tightly encloses the porous medium, such that the flow through the gap between them is negligible. At the same time, the gap is still wide enough that it allows free motion of the porous sphere at a given moment and that there are no direct mechanical forces across the gap (Fig.~\ref{fig:noslip}(a)). The rationale behind this assumption is that over longer times, the sphere, representing for example a microtubule aster (see Discussion), is dynamic and grows to fill the widening gap on one side and retracts on the other \citep{Tanimoto.Minc2016,DeSimone.Gonczy2018}. This growth and shrinkage take place through filament polymerisation and depolymerisation and therefore do not directly contribute to the force balance, but they maintain the system in a state where the assumption of a narrow gap remains valid. 

As with open boundaries, we first solve the problem for a point force and then for a pair of forces. Unlike in the previous cases, the force balance can be understood directly, without solving an auxiliary problem. 
For a point force in a fixed medium, it follows from considerations for confined spaces in Sec.~\ref{sec:confined} that the total force on the medium vanishes, $\bm{F}_\text{M}=\bm{0}$. Consequently, all the force is transmitted to the walls, so $\bm{F}_\text{W}=\bm{F}$. The result holds for any kind of boundary condition (no-slip or slip), as long as the normal velocity at the boundary is zero. 

In case of a pair of forces acting between the fluid and the medium (Fig.~\ref{fig:noslip}(b)), the problem is solved as superposition of the solution for the force $\bm{F}$ on the fluid while the medium is fixed (see above) and then for the force $\bm{F'}_\text{M}$ on the medium. The total force on the porous material is then $\bm{F}'_\text{M}=-\bm{F}$. In the second step, we need the solution for the problem where the medium is free to move and subject to the force $\bm{F}'_\text{M}$.

The problem where the medium is moving inside the tightly enclosing confinement without any other flows can be directly solved because the velocity inside the medium in this setting is spatially uniform (zero in the laboratory frame). The resulting drag coefficient is given by the volume of the sphere, multiplied by the drag density $\eta\alpha^2$, and reads
\begin{equation}
    \Gamma=\frac 4 3 \pi R^3 \eta \alpha^2\,.
\end{equation}
It follows that the velocity of the porous medium is 
\begin{equation}
\label{eq:vprime_enclosed}
    \bm{V}'=\frac{\bm{F}_\text{M}'}{\Gamma}=-\frac{\bm{F}}{6\pi \eta R} \cdot \frac 9 {2 (\alpha R)^2}\,.
\end{equation}
The solution of the full problem is then simply the superposition of the flow induced by a point force $\bm{F}$ in a fixed medium (see Appendix~\ref{appendix:d2} for an explicit derivation) and the medium moving with velocity $\bm{V}'$. Two examples of the solution are shown in Fig.~\ref{fig:noslip}(c,d). 
A comparison between Eq.~\eqref{eq:vprime_enclosed} and Eq.~\eqref{eq:vprime_open} shows that the velocity in confinement is always larger than in open fluid and that they only become equal if the force acts at the exterior boundary, $h=R$. Even though the confinement increases the drag coefficient, it increases the force $\bm{F}_\text{M}'$ even more.

\begin{figure}
\centering
\includegraphics[width=\textwidth]{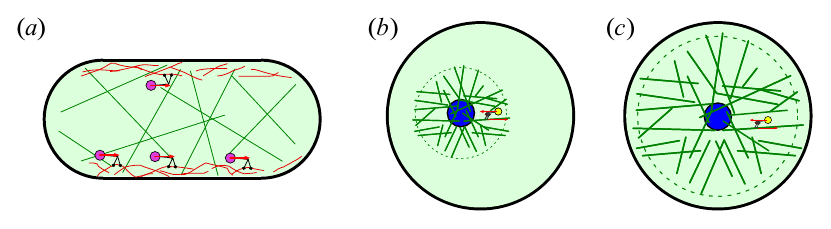}
\caption{(a)~Cytoplasmic streaming: molecular motors pull cargoes and exert forces (red arrows) on the cytoplasmic fluid. (b)~Dynein motors move vesicles along microtubules forming an aster, thereby pulling the pronucleus towards the centre of the cell. If the aster is small compared to the cell, it can be treated as a porous sphere immersed in an infinite fluid domain. (c)~Once the aster fills the volume of the cell, it becomes described by a porous sphere with no-slip boundaries. Because microtubules dynamically grow and shrink, their contact with the membrane does not affect the motion of the aster.}
\label{fig:applications}
\end{figure}

\section{Discussion}

The force balance in Brinkman fluids states that any external force is counterbalanced by the force on the porous material, forces on the walls and long-range stress (pressure). The distribution between these contributions takes different forms in different types of confinement. By applying the Lorentz reciprocal theorem, we could derive the integral forces in the setups we studied more easily than in previous explicit solutions for Newtonian fluids. We expect that the reciprocal theorem will also facilitate the solution of many problems involving 
Brinkman fluids, such as the hydrodynamic interaction between extended objects and point particles \citep{daddi23axisymmetric}. Although the problems discussed here all involve point forces acting on the fluid, or between the fluid and the medium, the solutions can also be expanded to other sources of flow, such as electroosmosis \citep{Coelho.Adler1996} or diffusioosmosis \citep{Keh2016}. In terms of force balance, these flows will have similar properties as collections of force pairs between the fluid and the medium, discussed in Sects.~\ref{sec:2.4} and \ref{sec:2.5}.

A natural application of these theoretical results are cytoplasmic flows in biological cells. The cytoplasm is often described with poroelastic models, where the cytosol flows through the cytoskeleton, which forms a porous, elastic mesh \citep{Moeendarbary.Charras2013,Mogilner.Manhart2018,Moradi.Nazockdast2022}. If we neglect the deformations of the cytoskeleton, and also its inhomogeneity and anisotropy, the Brinkman model becomes a good approximation for the cytoplasmic flows. 

A biological example closely related to the idealised cylindrical pipe studied in Sect.~\ref{sec:pipe} is the cytoplasmic streaming, a  flow with velocities up to $100\,\rm \mu m/s$ in some plant cells \citep{VANDEMEENT.GOLDSTEIN2010,Goldstein.vandeMeent2008}. The flow is driven by vesicles pulled by motors from the myosin-XI class. Similar cortical flows also appear in \textit{C.~elegans} zygotes, driven by kinesin-1 motors running along microtubules with polar orientation \citep{Kimura.Kimura2017} and in \textit{Drosophila} oocytes \citep{Ganguly.Goldstein2012}.

The spherical geometry, discussed in Sections~\ref{sec:2.4} and~\ref{sec:2.5}, finds its realization in the positioning of the male pronucleus in fertilised eggs \citep{Meaders.Burgess2020}, e.g.\ in \textit{C.~elegans} \citep{Shinar.Shelley2011,DeSimone.Gonczy2018} or sea urchin \citep{Tanimoto.Minc2016,Tanimoto.Minc2018}, which has also been reproduced in \textit{Xenopus} egg extracts \citep{Reinsch.Karsenti1997}. The pronucleus first grows a sperm aster consisting of polarised microtubules (Fig.~\ref{fig:applications}(b)). Dynein motors then induce a stream of cargo particles (such as endosomes, lysosomes or yolk granules) moving from the periphery to the centre of the aster. Because the microtubules grow shorter at the side facing the cell boundary, the resulting forces pull the pronucleus towards the centre of the cell and the female pronucleus. Although classical biophysical models describe the dynamics of the aster and the cytoplasm in a high level of detail \citep{Shinar.Shelley2011}, they treat the cytoplasm as a Newtonian fluid and do not consider the local hydrodynamic drag as it moves through the microtubule aster. The forces exerted by dynein motors on the microtubules are therefore used to pull the nucleus to the full extent. Later computational models coupled the cytoplasmic flow to the mechanics of cytoskeletal filaments \citep{Nazockdast.Shelley2017} and indeed showed that the aster behaves as a porous medium where the microtubules contribute the major part to its hydrodynamic drag. Our aim here is to use the simplified model to analytically understand the basic laws of force transduction in such structures.

In the continuum model we inevitable introduced a number of simplifications. 
In earlier simulations \citep{Nazockdast.Shelley2017} and experiments \citep{Tanimoto.Minc2018} the bending of microtubules under force did not play a major role and we therefore conclude that the assumption that the flows do not cause significant deformations of the network is well justified. We treat the cytoplasm as a Newtonian fluid, consistent with experiments showing deviations only on time scales shorter than a second \citep{Ganguly.Goldstein2012}. The polar orientation of microtubules in an aster at first contradicts the assumption of isotropy, but quantitative microscopic models of filamentous porous materials show that the anisotropy of permeability in such media is very small \citep{Wang2001}. There is also the assumption of uniform permeability, where we neglect the fact that the microtubules are denser at the centre. We further disregard any effects of inhomogeneity, i.e., any additional effect that could arise because the vesicles are moving along microtubules and not uniformly through the medium. While these details deserve a closer consideration, we expect that they will not significantly affect the main results.

To determine the velocity of the microtubule aster, we need to distinguish two situations. Initially, the aster is small and can be treated as a porous sphere in unconfined fluid, as studied in Sec.~\ref{sec:2.4}. Later, the aster fills most of the volume of the cell, while the dynamic nature of the microtubules still allows it to move inside it. The situation corresponds to the porous sphere with no-slip confinement (Sec.~\ref{sec:2.5}).

For a small aster, the effective force of the motors pulling on it is reduced by the factor $|F'_M/F|$, evaluated from Eq.~\eqref{eq:force-sphere} and averaged over the volume of the aster. The factor is 1 in the limit $\alpha\to 0$, as expected in a Newtonian fluid and assumed in \cite{Shinar.Shelley2011}. In a denser material, with $\alpha R \gg 1$, the efficiency of force generation is increasingly attenuated, as a large fraction of the force exerted by the vesicle on the fluid is transmitted back to the porous material. The attenuation, seen as the ratio between the solid and dashed lines in Fig.~\ref{fig:sphere}(d)~is stronger at the centre of the aster than at the periphery. 

Once the aster extends over the volume of the cell, the force transmission becomes 100\% efficient. The counter-intuitive consequence is that a dynein motor pulling a microtubule against a small vesicle produces as much force as if it was anchored to the membrane -- another common mechanism of centering \citep{Wu.Needleman2024}. We note, however, that this measure of efficiency is defined with respect to forces. Energetically, an anchored motor would still be more efficient because it moves at a smaller velocity. Furthermore, although its drag coefficient increases significantly as it extends over the whole volume of the cell, the aster still moves faster than in an unconfined fluid. 

The application of hydrodynamic models has decisively contributed to our understanding of many cellular processes \citep{Needleman.Shelley2019}. Whereas computational models are able to reproduce and interpret the experimental observations in great detail, we have shown that many fundamental aspects of force conservation and force transduction can be understood with idealised, analytically solvable, fluid-dynamic models. In addition to the examples shown here, we expect  the approach to be applicable to a wide range of hydrodynamic processes in cell biology which are currently being discovered at a rapid pace.

\vspace{0.5cm}

\textbf{Funding.}
This work was supported by the Slovenian Research and Innovation Agency (A.V., grant number P1-0099).
This investigation was initiated during A.V.’s research visit as part of the REF 2029 Visitors Programme, hosted by the STEM Faculty at the Open University in Milton Keynes.

\vspace{0.25cm}

\textbf{Declaration of interests.} The authors declare no conflicts of interest.

\appendix
\section{Lorentz reciprocal theorem for Brinkman fluids}
\label{appendix:lrt}

The Lorentz reciprocal theorem is a property of the Stokes flow that provides an integral identity connecting two different solutions of the Stokes equation: the main and the auxiliary problem \citep{masoud2019}. By using a known exact solution as auxiliary problem, integral quantities such as force on an object can be determined without explicitly solving the main problem. The reciprocal theorem can be generalized to flows described by the Brinkman equation. For 2D Brinkman flows, this was demonstrated by \cite{Bet.Samin2018}. The Brinkman fluid can also be seen as a special case of the poroelastic model, for which a reciprocal theorem has been derived recently \citep{Moradi.Nazockdast2024}. In the following, we recapitulate the derivation for Brinkman flows in 3D.
The derivation allows a spatially inhomogeneous parameter $\alpha$ and therefore holds, for example, for spaces partly consisting of the porous medium and partly of the ordinary fluid of the same viscosity. 

We consider two solutions (main and auxiliary) of the Brinkman equation (Eq.~\eqref{eq:brinkman_sigma}) with different force densities,
\begin{equation}
\nabla\cdot \boldsymbol{\sigma}-\eta\alpha^2\bm{v}+\bm{f}=\bm{0} \, , \qquad \nabla \cdot \boldsymbol{\hat \sigma}-\eta\alpha^2\bm{\hat v}+\bm{\hat f}=\bm{0}
\end{equation}
where the stress tensor of each flow is given by 
\begin{equation}
\boldsymbol{\sigma}=-p \bm{I}+ 2 \eta \bm{E} \, ,
\qquad
\boldsymbol{\hat \sigma}=-\hat p \bm{I}+ 2\eta \hat{\bm E}
\end{equation}
with $\bm{E}=\frac12 (\nabla \bm{v} + (\nabla \bm{v})^T)$.
We start with the identity
\begin{equation}
    \nabla \cdot (\bm{\hat\sigma}\cdot \bm{v})=(\nabla \cdot \bm{\hat \sigma})\cdot \bm{v}+\bm{\hat \sigma} \bm{:} \nabla \bm{v} = \eta \alpha^2 \bm{\hat v} \cdot \bm{v} -\bm{\hat f}\cdot \bm{v} -\hat p \nabla \cdot \bm{\hat v} + 2\eta \hat{\bm E}\bm{:}\bm{E}\,.
\end{equation}
The third term in the right-hand-side expression vanishes because $\nabla \cdot \bm{\hat v}=0$. By subtracting the equivalent identity for $\nabla \cdot (\bm{\sigma} \cdot \bm{\hat v})$, we obtain
\begin{equation}
    \nabla \cdot (\bm{\hat\sigma}\cdot \bm{v})-  \nabla \cdot (\bm{\sigma} \cdot \bm{\hat v}) = -\bm{\hat f}\cdot \bm{v} + \bm{f}\cdot \bm{\hat v}\,.
\end{equation}
After integrating over volume $\mathcal V$ and applying the divergence theorem, the reciprocal theorem follows
\begin{equation}
    \oint_{\partial \mathcal V} \bm{v} \cdot \bm{\hat \sigma} \cdot \bm{n} \, \dd S + \int_{\mathcal V}  \bm{v}\cdot \bm{\hat f}  \, \dd V =
\oint_{\partial \mathcal V}\bm{\hat v} \cdot \bm{\sigma} \cdot  \bm{n} \, \dd S + \int_{\mathcal V}  \bm{\hat v} \cdot \bm{f} \, \dd V \, , 
\end{equation}
with \(\bm{n}\) representing a unit surface normal vector pointing outward.

\section{Axisymmetric point force in a porous tube}
\label{appendix:pipe}

To determine the solution for a point-force singularity acting along the axis of a porous tube, we employ the image solution technique~\citep{happel12}. 
The core idea is to express the solution as a superposition of the free-space solution (in the absence of confinement) and an image solution that is force-free in the interior and together with the free-space solution satisfies the boundary conditions. We use cylindrical coordinates and represent the solution using Fourier-Bessel integrals.

By taking the divergence of the Brinkman equation~\eqref{eq:brinkman}, we obtain a Poisson equation for the pressure:
\begin{equation}
    \Delta p =
    F \, \frac{\partial }{\partial z} \, \delta( \bm{r} ) \, .
\end{equation}
In an infinite Brinkman fluid medium, its solution (equivalent to the potential of an electrostatic dipole) in Fourier-Bessel space is given by 
\begin{equation}
	p^\infty = \frac{F}{2\pi^2} \int_0^\infty 
	k K_0(k\rho) \sin(kz) \, \mathrm{d} z \, .
\end{equation}

For a given pressure, the velocity field  $v_z^\infty$ is determined by the Helmholtz equation 
\begin{equation}
    \left( \Delta - \alpha^2 \right) v_z^\infty = \frac{\partial p^\infty}{\partial z} - F \delta(\bm{r}) \,.
\end{equation}
It is solved by
\begin{equation}
    v_z^\infty =
	\frac{F}{2\pi^2 \eta \alpha^2} \int_0^\infty
	\left(  - k^2 K_0(k\rho) + q^2 K_0({q}\rho)\right) \cos(kz) \, \mathrm{d} k \, ,
\end{equation}
with $q = \sqrt {k^2 + \alpha^2}$.
The radial component $v_\rho^\infty$ can then be determined from the incompressibility equation as
\begin{equation}
    v_\rho^\infty =
	\frac{F}{2\pi^2 \eta \alpha^2} \int_0^\infty
	k \left( k  K_1(k\rho) - {q}  K_1({q}\rho) \right)
	\sin(kz) \, \mathrm{d} k \, .
\end{equation}
By evaluating the above integrals, we obtain the known fundamental solution of the Brinkman equation \citep{howells1974drag}
\begin{equation} 
p^\infty = \frac{F}{4\pi} \frac{z}{s^3}\,, \quad
		v_z^\infty = \frac{F}{8\pi\eta} \left( \frac{B_1(\beta)}{s} + B_2(\beta) \, \frac{z^2}{s^3}\right)   \, , \quad
  v_\rho^\infty = \frac{F }{8\pi\eta} \,  \frac{B_2(\beta) \rho z}{s^3} 
		 \, .
	\label{eq:brinkmanlet}
\end{equation}
We have defined the abbreviation $\beta = \alpha s$ with \( s = \left( \rho^2 + z^2 \right)^\frac{1}{2} \) denoting the distance from the singularity position.
The functions \( B_1 \) and \( B_2 \)  are given by
\begin{equation}
		B_1(\beta) = 2e^{-\beta} 
		\left( 1 + \frac{1}{\beta} + \frac{1}{\beta^2} \right) - \frac{2}{\beta^2} \, , \quad
		B_2(\beta) = \frac{6}{\beta^2} - 2 e^{-\beta} \left( 1 + \frac{3}{\beta} + \frac{3}{\beta^2} \right) \, ,
\end{equation}
In the limit \(\alpha \to 0\), \(B_1 = B_2 = 1\), which simplifies Eqs.~\eqref{eq:brinkmanlet} to the classical Stokeslet.

To determine the image solution, we start with an arbitrary axisymmetric solution of the homogeneous equation \( (\Delta -\alpha^2)\bm{v}=\bm{\nabla} p \). Again, by taking its divergence, we obtain $\Delta p=0$. A general axisymmetric solution of the Laplace equation can be written as
\begin{equation}
	p^* = -\frac{F}{2\pi^2} \int_0^\infty A(k) I_0(k\rho) \sin(kz)
	\, \mathrm{d} k \, ,
\end{equation}
with an unknown function $A(k)$. A general solution of the Brinkman equation follows as a superposition of the particular solution $\bm{v}=-\alpha^{-2} \bm{\nabla} p$ and a general solution of the homogeneous equation \( (\Delta - \alpha^2)v_z =0\):
\begin{eqnarray}
v_z^* &=& \frac{F}{2\pi^2\eta \alpha^2} 
	\int_0^\infty \left( k A(k)I_0(k\rho) + q B(k) I_0(q\rho) \right) \cos(kz) \, \mathrm{d} k \, , \label{eq:vzStar}\\
	v_\rho^* &=& \frac{F}{2\pi^2\eta \alpha^2} 
	\int_0^\infty k \left( A(k)I_1(k\rho) + B(k) I_1(q\rho) \right) \sin(kz) \, \mathrm{d} k \, , \label{eq:vrStar} 
\end{eqnarray}
with another arbitrary function $B(k)$. Again, the radial function was determined such that it satisfies $\bm{\nabla}\cdot \bm{v}=0$.

By enforcing no-slip boundary conditions at \( \rho = R \), which correspond to vanishing radial and axial velocities, the functions \( A \) and \( B \) are determined as
\begin{equation}
    A = \frac{\Phi(k,q) }{ R D}  \, , \qquad
    B = \frac{\Phi(q,k) }{ R D} \, \, , 
\end{equation}
with
\begin{equation}
    \Phi(x,y) = 
    xR \left( y  K_1(xR)I_0(yR) + x  K_0(xR)I_1(yR) \right) - y \, ,
\end{equation}
and denominator
\begin{equation}
    D = kI_0(kR)I_1(qR) - qI_0(qR)I_1(kR) \, .
\end{equation}
Examples of the resulting flow fields are shown in Fig.~\ref{fig:tube}(d). For small values of $\alpha R$, the flow contains viscous toroidal eddies, which were already studied by \cite{Blake1979} in a Stokes fluid. 
However, the eddies vanish when the value of $\alpha R$ exceeds approximately 4.7.

The hydrodynamic force exerted on the inner wall of the tube is calculated by integrating the hydrodynamic viscous stress tensor over the cylindrical wall
\begin{equation}
	F_\mathrm{W} = -2\pi R \left. \int_{-\infty}^\infty \sigma_{\rho z} \right|_{\rho = R} \, \mathrm{d}z \, ,
\end{equation}
with $\sigma_{\rho z} = \eta \left( \partial_\rho v_z + \partial_z v_\rho \right)$.
Because \( v_\rho \) vanishes at $\rho=R$, the force involves the contribution from \( \eta \partial_z v_\rho \) only. 
By first integrating with respect to~\( z \) and then with respect to \( k \), we obtain
\begin{equation}
	F_\mathrm{W} = -\frac{2I_1(\alpha R)/(\alpha R)-1}
 { I_2(\alpha R) } \, F \, , 
\end{equation}
in agreement with Eq.~\eqref{eq:FW_pipe} when setting $h=0$.
We note that \( F_\mathrm{W} \) varies between~\(-F\) and~\(0\).

The force due to the pressure at both ends of the tube can be evaluated as
\begin{equation}
	\frac{F_\mathrm{P}}{2} = \pm \pi R^2 \lim_{z \to \pm \infty}	 p  \, .
\end{equation}
Because $p^\infty$ decays with $\left| z \right|^{-2}$, its contribution to the integral is zero.  To account for the contribution from the image solution $p^*$, we use a variable substitution \( u = kz \) and take the limit \( z \to \pm\infty \). 
Note that \( p^* \) becomes independent of \( \rho \) as \( z \) approaches infinity.
We obtain
\begin{equation}
	F_\mathrm{P} = 
 \frac{  I_0(\alpha R)-1  }
 { I_2(\alpha R) } \, F  \, ,
\end{equation}
which varies between $F$ and $2 F$.
This equation aligns with the pressure force derived from Eq.~\eqref{eq:Delp_tube} when setting \( h = 0 \).

It can readily be checked that
\begin{equation}
	F = F_\mathrm{W} + F_\mathrm{P} \, ,
\end{equation}
so that part of the force is transmitted to the wall and the remainder manifests as pressure at both ends of the tube. The total force the fluid exerts on the porous material is therefore zero.

\section{Parallel point force in a porous medium near a wall}
\label{appendix:wall}

Next, we examine the case of a point force acting parallel to an infinitely extended no-slip wall in the \(xy\)-plane. 
The point force is positioned at a distance~\( h \) above the wall.
Without loss of generality, we assume the point force is directed along the \(x\)-axis. To satisfy the no-slip boundary conditions at the wall, we employ the image solution technique and represent the solution as the superposition of a free-space Brinkmanlet and an image that is force-free in the fluid domain.

The problem can be solved using a 2D Fourier transform technique.
Accordingly, in 2D Fourier space, the flow variables are expressed using \( k_x = k \cos \phi \) and \( k_y = k \sin \phi \), where \( k \) represents the wavenumber. Note that the \( z \)-coordinate is not transformed.
The 2D Fourier-transformed Brinkman equations read
\begin{eqnarray}
    -i k_x \widetilde{p} + \eta \left( \partial_{zz} \widetilde{v}_x -q^2 \widetilde{v}_x \right)  &=& - \widetilde{F}_x \, \delta(z-h) \, , \label{eq:X} \\
    -i k_y \widetilde{p} + \eta \left( \partial_{zz} \widetilde{v}_y -q^2 \widetilde{v}_y \right)  &=& 0 \, , \label{eq:Y} \\
    -\partial_z \widetilde{p} + \eta \left( \partial_{zz} \widetilde{v}_z -q^2 \widetilde{v}_z \right)  &=& 0 \, , \label{eq:Z}
\end{eqnarray}
where, again, $q=\sqrt{ k^2+\alpha^2}$.
The incompressibility condition in 2D Fourier space is expressed as
\begin{equation}
\label{eq:plane-conitnuity}
    i k_x \widetilde{v}_x + i k_y \widetilde{v}_y + \partial_z \widetilde{v}_z = 0 \, .
\end{equation}

We use the method introduced by \cite{bickel06, bickel07} and define the longitudinal and transverse components of the velocity field as 
\begin{equation}
    \widetilde{v}_l = \widetilde{v}_x \cos\phi + \widetilde{v}_y \sin\phi \, , \qquad
    \widetilde{v}_t = \widetilde{v}_x \sin\phi - \widetilde{v}_y \cos\phi \, ,
\end{equation}
and similarly for the longitudinal and transverse components of the force \citep{daddi16c, daddi2018brownian}.

The continuity equation (\eqref{eq:plane-conitnuity}) establishes a direct relationship between the longitudinal and normal components of the velocity through
\begin{equation}
    \widetilde{v}_l = \frac{i}{k} \, \partial_z \widetilde{v}_z \, . \label{eq:vl}
\end{equation}

From Eqs.~\eqref{eq:X} and~\eqref{eq:Y} we find that the transverse component of the velocity field is governed by the following second-order differential equation
\begin{equation}
    \partial_{zz} \widetilde{v}_t - q^2 \widetilde{v}_t =
    -\frac{ \widetilde{F}_t }{\eta} \, \delta(z-h) \, . \label{eq:vt}
\end{equation}

By taking the partial derivatives of Eqs.~\eqref{eq:X} and~\eqref{eq:Y} with respect to~$z$, and using Eq.~\eqref{eq:plane-conitnuity} to eliminate $\tilde v_x$ and $\tilde v_y$ and Eq.~\eqref{eq:Z} to eliminate the pressure, we obtain the following fourth-order differential equation for the $z$-component of the velocity field
\begin{equation}
    \partial_{zzzz} \widetilde{v}_z - \left( k^2+q^2\right) \partial_{zz} \widetilde{v}_z + k^2 q^2 \widetilde{v}_z = 
    \frac{i k \widetilde{F}_l }{\eta} \, \delta'(z-h) \, . \label{eq:vz}
\end{equation}

Considering the regularity conditions at infinity, only the terms with decaying exponentials should be retained.
In a bulk fluid medium, the solution of Eqs.~\eqref{eq:vl} through~\eqref{eq:vz} is~\citep{daddi16}
\begin{eqnarray}
   \widetilde{v}_t^\infty &=& \frac{ \widetilde{F}_t }{2\eta q} \, e^{-q|z-h|} \, , \\
   \widetilde{v}_z^\infty &=& 
   \frac{ik \widetilde{F}_l }{2\eta \alpha^2} \,
   \operatorname{sgn} (z-h)
   \left( e^{-q|z-h|} - e^{-k|z-h|} \right) \, , \\
   \widetilde{v}_l^\infty &=& 
   \frac{ \widetilde{F}_l }{2\eta \alpha^2}
   \left( q e^{-q|z-h|} - k e^{-kz} \right) \, ,
\end{eqnarray}
with $\operatorname{sgn} (x)=x/|x|$ denoting the sign function.
The corresponding pressure follows from Eq.~\eqref{eq:Z} 
\begin{equation}
    \widetilde{p}^\infty = -\frac{i \widetilde{F}_l }{2} \, e^{-k |z-h|} \, .
\end{equation}

The image solution can be represented as a general solution of the homogeneous equations analogue to Eqs.~\eqref{eq:vt} and~\eqref{eq:vz}
\begin{equation}
    \widetilde{v}_t^* = \frac{ \widetilde{F}_t }{2\eta} \, A e^{-qz} \, , \qquad
    \widetilde{v}_z^* = \frac{ i \widetilde{F}_l }{2\eta \alpha^2} \left( B e^{-kz} + C e^{-qz} \right) \, , 
\end{equation}
where \( A \), \( B \), and \( C \) are unknown wavenumber-dependent functions. 
In addition, $\widetilde{v}_l^*$ can be determined from Eq.~\eqref{eq:vl} as
\begin{equation}
    \widetilde{v}_l^* = \frac{ \widetilde{F}_l }{2\eta \alpha^2} \left( B e^{-kz} + \frac{q}{k} \, C e^{-qz} \right) \, .
\end{equation}
The corresponding solution for the pressure is 
\begin{equation}
    \widetilde{p}^* = \frac{ i\widetilde{F}_l }{2k} \, B e^{-kz} \, .
\end{equation}

The functions $A$, $B$ and $C$ are obtained from the conditions that $\tilde v_t^\infty + \tilde v_t^* =0$, $\tilde v_l^\infty + \tilde v_l^* =0$ and
$\tilde v_z^\infty + \tilde v_z^* =0$ at $z=0$ as
\begin{equation}
    A = -\frac{e^{-qh} }{q} \, , \qquad
    B = k\, \frac{2q e^{-qh} - (k+q)e^{-kh}}{q-k}  \, , \qquad
    C = k\,\frac{2k e^{-kh} - (k+q)e^{-qh}}{q-k}  \, . \notag
\end{equation}

To calculate the force exerted on the wall, we integrate the stress \(\sigma_{xz} = \eta \left( \partial_x v_z + \partial_z v_x \right)\) over the surface of the wall. Because $v_z=0$ at the boundary, the first term vanishes and force can be calculated as
\begin{equation}
    F_\mathrm{W} 
    = \frac{\eta}{2\pi} \lim_{k\to 0}
    \int_0^{2\pi} \partial_z \widetilde{v}_x \big|_{z=0} \, \mathrm{d}\phi
    = F e^{-\alpha h} \, ,
\end{equation}
in agreement with Eq.~\eqref{eq:force-wall}.

The far-field pressure is obtained by evaluating $\tilde p = \tilde p^\infty+ \tilde p^*$ in the limit of small $k$ and transforming the result back to real space, which gives 
\begin{equation}
\label{eq:plane-p}
p=\frac {F x}{2\pi r^3} \left( 1 - e^{-\alpha h} \right)\,.
\end{equation}

\section{Axisymmetric point force inside a porous sphere}
\label{appendix:sphere}

We employ the image solution technique to find the solution for a point force acting at position $h \in [0, R]$ along the $z$-axis. The radial and azimuthal velocities can be expressed with a stream function $\psi$ as 
\begin{equation}
\label{eq:v-from-psi}
    v_r = \frac{1}{r^2 \sin\theta} \frac{ \partial \psi }{ \partial \theta } \, , \qquad
    v_\theta = -\frac{1}{r \sin\theta} \frac{ \partial \psi }{ \partial r } \, .
\end{equation}
The Brinkman equation in terms
of the stream function then reads \citep{Pop.Ingham1996}
\begin{equation}
    \mathcal{D}^2 \left( \mathcal{D}^2 - \alpha^2 \right) \psi = 0 \, , 
\end{equation}
with the differential operator 
\begin{equation}
    \mathcal{D}^2 = \frac{ \partial^2 }{\partial r^2} + \frac{ \sin\theta }{r^2}
    \frac{\partial}{\partial \theta}
    \left( \frac{1}{\sin\theta} \frac{\partial}{\partial \theta} \right) \, .
\end{equation}
We express the solution of the Brinkman equation inside the porous sphere as a superposition of the free space solution around the point force and a force-free image solution:
\begin{equation}
    \psi^{(i)} = \psi^\infty + \psi^* \, .
\end{equation}
The corresponding solution of the Stokes flow outside the porous sphere is denoted by \(\psi^{(o)}\).

By decomposing the stream function as \(\psi = \psi^{(1)} + \psi^{(2)}\), where \(\psi^{(1)}\) satisfies \( \mathcal{D}^2 \psi^{(1)} = 0\) and \(\psi^{(2)}\) satisfies \(\left( \mathcal{D}^2 - \alpha^2 \right) \psi^{(2)} = 0\), the stream functions can be decomposed into angular modes as \citep{palaniappan2014some, nganguia2018squirming}
\begin{equation}
    \label{eq:psi-expansion}
	\psi = \frac{F}{8\pi\eta} \sum_{n=1}^\infty
	\psi_n(r) P_n^1 (\cos\theta) \sin \theta \, ,
\end{equation}
where $P_n^1 (\cos\theta) = \mathrm{d}P_n(\cos\theta) / \mathrm{d}\theta$ are representing the associated Legendre polynomials of the first order.
Likewise, we express the pressure field as
\begin{equation}
	p = \frac{F}{8\pi} \sum_{n=1}^\infty
	p_n(r) P_n (\cos\theta) \, .
\end{equation}

By expressing Eqs.~\eqref{eq:brinkmanlet} in spherical coordinates, the free-space stream function for a Brinkman fluid medium can be derived as
\begin{equation}
    \psi^\infty = \frac{F}{4\pi\eta \alpha^2}
    \frac{r^2}{s^3} 
    \left( 1 - (1+\alpha s) e^{-\alpha s} \right) \sin^2 \theta \, , 
\end{equation}
with $s = \left( r^2+h^2-2hr\cos\theta \right)^\frac{1}{2}$ representing the distance from the singularity position. For simplicity, we rescaled all lengths by the sphere radius $R$. The above solution can be expanded according to Eq.~\eqref{eq:psi-expansion} by using the orthogonality properties of Legendre polynomials. The radial part of each term is derived as
\begin{subequations}
\begin{eqnarray}
    \psi_n^\infty &=& \frac{2}{\alpha^2 h}
	\left( (2n+1) \left( \frac{r}{h} \right)^\frac{1}{2} I_{n+\frac{1}{2}} (\alpha h) K_{n+\frac{1}{2}} (\alpha r)  - \left( 
	\frac{h}{r} \right)^n \right) \quad \text{for} \; r>h \, , \\
 \psi_n^\infty &=& 
    \frac{2}{\alpha^2 h}
	\left( (2n+1) \left( \frac{r}{h} \right)^\frac{1}{2} K_{n+\frac{1}{2}} (\alpha h) I_{n+\frac{1}{2}} (\alpha r)  - \left( 
	\frac{r}{h} \right)^{n+1} \right) \quad \text{for} \; r<h\qquad
\end{eqnarray}
\end{subequations}
and the corresponding pressure field is 
\begin{subequations}
\begin{eqnarray}
    p_n^\infty &=& \frac{2n}{r^2} \left( \frac{h}{r} \right)^{n-1} 
    \quad \text{for} \; r>h \, ,\\
    p_n^\infty &=& -\frac{2(n+1)}{h^2} \left( \frac{r}{h} \right)^n \quad \text{for} \; r<h\, . 
\end{eqnarray}
\end{subequations}

For the image solution in the Brinkman fluid inside the sphere, we use the general solution \citep{nganguia2018squirming}
\begin{equation}
        \psi_n^* = \frac{1}{\alpha^2} \left( A_n r^{n+1} + B_n r^\frac{1}{2} I_{n+\frac{1}{2}}(\alpha r) \right) , 
\end{equation}
with the pressure
\begin{equation}
        p_n^* = (n+1) A_n r^n \, .
\end{equation}

Outside the sphere, a general axisymmetric solution of the Stokes equation is typically expressed in terms of Gegenbauer polynomials \citep{happel12, kim05}. Transformed to the basis defined in Eq.~\eqref{eq:psi-expansion}, it can be written as 
\begin{equation}
  \psi_n^{(o)} = \frac{a_n}{r^n} + \frac{b_n}{r^{n-2} } \, ,
\end{equation}
with
\begin{equation}
    p_n^{(o)} = -2n(2n-1) \, \frac{b_n}{ r^{n+1} } \,.
\end{equation}      
The components of the fluid velocity, which follow from the stream function via Eq.~\eqref{eq:v-from-psi}, can be cast in the form
\begin{eqnarray}
    v_r = \frac{F}{8\pi\eta} \sum_{n=1}^\infty R_n (r) P_n(\cos\theta) \, , \qquad
    v_\theta = \frac{F}{8\pi\eta} \sum_{n=1}^\infty T_n (r) P_n^1(\cos\theta) \, .
\end{eqnarray}

For the free-space components, we obtain
\begin{subequations}
    \begin{eqnarray}
    R_n^\infty &=& \frac{2n(n+1)}{\alpha^2 h r^2} 
    \left( 
    -(2n+1) \left( \frac{r}{h} \right)^\frac{1}{2}
    I_{n+\frac{1}{2}} (\alpha h)
    K_{n+\frac{1}{2}}(\alpha r)  + \left(\frac{h}{r} \right)^n 
    \right) ,\qquad \\
    T_n^\infty &=& \frac{2}{\alpha^2 h r^2}
    \left( (2n+1) \left( \frac{r}{h} \right)^\frac{1}{2}
    I_{n+\frac{1}{2}}(\alpha h) H_n(\alpha r)
    - n \left( \frac{h}{r} \right)^n
    \right) ,
\end{eqnarray}
\end{subequations}
where we have defined the abbreviation
\begin{equation}
    H_n (\alpha r) = \alpha r K_{n-\frac{1}{2}}(\alpha r) + n K_{n+\frac{1}{2}}(\alpha r) \, .
\end{equation}
For the image solution, we have
\begin{subequations}
    \begin{eqnarray}
    R_n^* &=& -\frac{n(n+1)}{\alpha^2 r^2}
    \left( A_n r^{n+1} + B_n r^\frac{1}{2} I_{n+\frac{1}{2}}(\alpha r) \right) , \\
    T_n^* &=& -\frac{1}{\alpha^2 r^2}
    \left( (n+1)A_n r^{n+1} + B_n r^\frac{1}{2} Q_n (\alpha r) \right) ,
\end{eqnarray}
\end{subequations}
with 
\begin{equation}
    Q_n (\alpha r) = \alpha r  I_{n-\frac{1}{2}}(\alpha r) - n I_{n+\frac{1}{2}}(\alpha r) \, .
\end{equation}
The solution of the Stokes flow outside the porous sphere is
\begin{subequations}
    \begin{eqnarray}
        R_n^{(o)} &=&  -n(n+1) \left( \frac{a_n}{r^{n+2}} + \frac{b_n}{r^n} \right) , \\
        T_n^{(o)} &=& \frac{n}{r^{n+2}} \, a_n + \frac{n-2}{r^n}\, b_n \, .
    \end{eqnarray}
\end{subequations}

The unknown sequences \( A_n \), \( B_n \), $a_n$, and~$b_n$ can be determined based on whether the porous sphere has open boundaries or no-slip confinement.

\subsection{Porous sphere with open boundaries}

For a porous sphere with open boundaries, it is necessary to satisfy the continuity of both the velocity and the viscous stress tensor at \( r = 1 \).
The coefficients within the porous sphere can be expressed in the form
\begin{equation}
    A_n = \frac{ 2n M_n }{D_n} \, , \qquad
    B_n = \frac{ 2(2n+1) N_n }{D_n} \, ,
\end{equation}
where
\begin{eqnarray}
    M_n &=&   (4n^2-1)I_{n+\frac{1}{2}} (\alpha h) -\alpha h^{n+\frac{1}{2}} \left( (2n-1)I_{n - \frac{1}{2}} (\alpha) + \alpha I_{n + \frac{1}{2}} (\alpha)\right) , \\
    N_n &=&
     n(2n-1)h^{n+\frac{1}{2}} 
    -\Lambda_n I_{n+\frac{1}{2}} (\alpha h) ,
\end{eqnarray}
with 
\begin{equation}
    \Lambda_n = 
    \alpha (n+1) \left( \alpha K_{n + \frac{1}{2}} (\alpha)-(2n-1)K_{n - \frac{1}{2}} (\alpha) \right)+n(4n^2-1)K_{n + \frac{1}{2}} (\alpha) 
    .
\end{equation}
The denominator is given by
\begin{equation}
    D_n = h^\frac{3}{2}
    \left( \alpha (n+1) \left( \alpha I_{n + \frac{1}{2}} (\alpha) + (2n-1)I_{n - \frac{1}{2}} (\alpha) \right) + n(4n^2-1) I_{n + \frac{1}{2}} (\alpha) \right) . \notag
\end{equation}

Outside the porous sphere, the coefficients can similarly be expressed in the form
\begin{equation}
    a_n = \frac{ (2n+1) S_n }{\alpha^2 D_n} \, , \qquad
    b_n = -\frac{ (2n+1) W_n }{D_n} ,
\end{equation}
where 
\begin{eqnarray}
    S_n &=&
     \alpha \left( \alpha(n-2) I_{n + \frac{1}{2}} (\alpha) - 2(2n-1)I_{n - \frac{1}{2}} (\alpha) \right) h^{n+\frac{1}{2}} 
    + \Delta_n I_{n+\frac{1}{2}} (\alpha h) , \qquad \\
    W_n &=&
     nI_{n + \frac{1}{2}} (\alpha) h^{n+\frac{1}{2}} + (n+1)I_{n+\frac{1}{2}} (\alpha h) ,
\end{eqnarray}
with 
\begin{equation}
    \Delta_n = 2(4n^2-1)+\alpha^2(n+1) \, .
\end{equation}
The streamlines of the resulting flow are shown in Fig.~\ref{fig:sphere}(e,f) for the values $h=0$ and $h=0.75$. 

The hydrodynamic force exerted on the background fluid can be obtained by surface integration 
\begin{equation}
    F_z = 2\pi
    \int_{r>1} \left( \sigma_{rr}^{(o)} \cos\theta - \sigma_{r\theta}^{(o)} \sin\theta \right) r^2 \sin\theta \, \mathrm{d} \theta \, ,
\end{equation}
leading to the same result as Eq.~\eqref{eq:force-sphere} in the main text derived using the reciprocal theorem.

\subsection{Porous sphere with no-slip confinement} 
\label{appendix:d2}

Considering no-slip boundary conditions, where the velocity is zero at \( r = 1 \), the series coefficients of the solution inside the porous sphere are determined as
\begin{equation}
    A_n = \frac{2\alpha C_n}{L_n} \, , \qquad
    B_n = \frac{2(2n+1)E_n}{L_n} \, , 
\end{equation}
where
\begin{eqnarray}
    C_n &=& 
    -(2n+1) \left( I_{n - \frac{1}{2}} (\alpha) K_{n + \frac{1}{2}} (\alpha) + I_{n + \frac{1}{2}} (\alpha) K_{n - \frac{1}{2}} (\alpha) \right) I_{n+\frac{1}{2}} (\alpha h) 
    \nonumber \\ &&+I_{n - \frac{1}{2}} (\alpha) h^{n+\frac{1}{2}}\, , \\
    E_n &=& \left( (2n+1) K_{n + \frac{1}{2}} (\alpha) + \alpha K_{n - \frac{1}{2}} (\alpha) \right) I_{n+\frac{1}{2}} (\alpha h) -h^{ n+\frac{1}{2} } \, .
\end{eqnarray}
The denominator is given by
\begin{equation}
    L_n = h^\frac{3}{2} 
    \left( \alpha I_{n - \frac{1}{2}} (\alpha) - (2n+1) I_{n + \frac{1}{2}} (\alpha) \right) \, .
\end{equation}
Two examples of the flow are shown in Fig.~\ref{fig:noslip}(c,d).

\bibliography{biblio}
 
\end{document}